# Hyperactivity in 103P/Hartley 2: Chunks from the sub-surface in Type IIa jet regions.

Michael J.S. Belton

*Belton Space Exploration Initiatives, 430 S. Randolph Way, Tucson, AZ 85716, USA*

*Emeritus Astronomer, Kitt Peak National Observatory, Tucson, AZ, 85719 USA*

Pages      32

Figures     5

Tables      1

Corresponding author:

Michael J.S. Belton

Belton Space Exploration Initiatives, LLC

430 S Randolph Way

Tucson, AZ 85716

Phone: 520-795-6286

Cell: 520-275-0066

Fax: 520-795-6286

Email: mbelton@dakotacom.net

**ABSTRACT**

We analyze the observed radial distribution of column densities of water-ice particulates embedded in the primary jet region ($J_1$) of 103P's inner coma at altitudes between 439 and 1967 m (Protopapa *et al.*, 2014, *Icarus* 238, 191-204) and determine the speed and acceleration of particles and their mass-flow within the filaments of the jet. This is done by applying a $CO_2$ driven (Type IIa) jet model proposed by Belton (2010, *Icarus* 210, 881-897). The model utilizes water-ice particles dislodged in the source regions of the jet filaments and accelerated by $CO_2$ to explain the radial distribution of water-ice particulates. We provide an explanation for the remarkably different radial distribution of refractory dust particles by hypothesizing that the majority of the dust originates directly from the nucleus surface in inter-filament regions of the jet complex and is accelerated by $H_2O$. Our model provides a mass-flow of water from the $J_1$ jet complex that is ~40 times greater than the constant speed sublimation model discussed by Protopapa *et al.* but is still too small to explain the hyperactivity of the comet. Speeds in the flow are increased by a factor up to ~20 over those found by Protopapa *et al.*

To account for the hyperactivity, most of the mass dislodged in the filament source regions must be in weakly accelerated large chunks that achieve only low speeds *en route* to the region of observation. These chunks soon leave the filamentary jet structure due to the rotation of the nucleus and do not contribute to the column densities observed at higher altitudes in the jet filaments. Employing the results of Kelley *et al.* (2015. *Icarus* 262, 187-189)) on total cross-section and mass-flow in the coma we find that large chunks with the same bulk properties as the nucleus can increase the active fraction of the comet by two or three times. With the exception that the chunks do not need to be "nearly pure water ice", these results support the hypothesis that hyperactivity in comet 103P/Hartley 2 is the result of super-volatiles that "…drag out chunks of nearly pure water ice that then sublime to provide a large fraction of the total $H_2O$ gaseous output of the comet (A'Hearn *et al*, 2011. *Science* 332, 1396-1400). We find that the largest chunks (effective radius ~4 m) are probably dislodged at an average rate less than 1 chunk / filament / 1- 3 rotation periods near perihelion and we estimate that the filamentary source regions, if 50 - 100m in diameter, could be excavated to a depth of 44 - 171m in a single perihelion passage. We note similarities with groups of active pits discovered on 67P (Vincent *et al.,* 2015. *Nature* 523, 63 – 66) and suggest that these features may have, in the past, supported Type IIa super-volatile jet outflows that are now essentially exhausted.

## 1. Introduction.

The primary purpose of this paper is to present an analysis of Protopapa *et al's* (2014) measurements of the radial distribution of water-ice bearing particulate column densities in a $CO_2$ rich jet complex located in the innermost coma (altitude range: 439 – 1967 m) of 103P/Hartley 2. We do this in terms of a mixture of fine refractory dust and water-ice particles, dislodged from the sub-surface by turbulent $CO_2$ effusion within the source regions of the filamentary structure of Type IIa jets (Belton, 2010)[1] and thence accelerated up the filaments. A secondary goal is to account for the hyperactivity[2] of the comet. This is done by recognizing that most of the dislodged mass resides in large chunks that leave the filaments source regions with sufficiently small acceleration so that, as a result of the rotation of the nucleus, they leave the filamentary structure and do not directly contribute to Protopapa *et al's* observations. They, nevertheless, persist on orbits in the inner coma subliming $H_2O$ and contribute to the hyperactivity.

There have been various previous suggestions made on the source of the hyperactivity. Lisse *et al.* (2009) suggested that icy particles in the comet's coma might be responsible. Later, in an overview of the scientific results of the EPOXI encounter, A'Hearn *et al.* (2011) suggested that super-volatiles, in this case $CO_2$, "…drag out chunks of nearly pure water ice, which then sublime to provide a large fraction of the total $H_2O$ gaseous output of the comet." A'Hearn *et al's* concept was tested observationally by Kelley *et al.* (2013, 2015) who found that the observed flux distribution of large coma particles, if assumed to be pure solid water-ice, could at most contribute ~0.2% of the comet's water production rate and suggested that material observed in the prominent $CO_2$ rich jets displayed by the comet might instead be the primary cause of the hyperactivity. Kelley *et al.* (2015) also considered the case of water-ice carrying dusty aggregates and showed that, if they had the properties of mini-comet nuclei, they would have to be an order of magnitude larger (up to 4 m in effective radius) than solid ice particles. However, they could potentially produce 30-200% of the observed $H_2O$ production rate thus providing an explanation for the hyperactivity. In their original paper this possibility was dismissed because it implied a coma mass greater than that of the nucleus (Kelley *et al.*, 2013). This mass estimate has now been revised downward by a factor of ~$10^3$ (Kelley *et al.*, 2015) to as low as 0.6% of $M_{nucleus}$. Nevertheless, because of the large and massive particles it implies, this case was not favored by Kelley *et al*. (2015) even though it apparently resolved the source of the comet's current hyperactivity.

---

[1] Belton (2010) proposed a taxonomy for active areas: Types I and II are persistent and Type III episodic. Type I are driven by sublimating $H_2O$; Type II by super-volatiles. Type IIa are associated with filamentary outflows.

[2] A comet is said to be hyperactive when its observed $H_2O$ production rate is larger than the rate at which water molecules can be released by sublimation from water-ice covering the entire surface of the nucleus. A few such comets are known including 103P/Hartley 2 that was encountered by the EPOXI spacecraft on November 4, 2010 at a distance of 1.06 AU from the sun (Lisse *et al.,* 2009; A'Hearn *et al.,* 2011). Comet 103P/Hartley 2 has an active fraction between 0.3 - 1 (Groussin *et al.,* 2010)

In a separate investigation, Protopapa *et al*. (2014) investigated the spectral character and radial distribution of water-ice carrying grains and dust in two jet regions of the comet ($J_1$ and $J_4$) in the infra-red (1 - 5μ). Their analysis was based on the sublimation of small (1 μm radius) particles moving at constant speed in the jets. They found an upper limit of $2 \times 10^{-4}$% for their contribution to the water production rate and concluded that an explanation of hyperactivity based on pure water-ice particles is "…still unable to solve the mystery about the source of water in Hartley 2's coma." Protopapa *et al*. did note that an acceleration of water-ice carrying aggregates in the altitude region of interest might provide an alternate explanation the observed radial distribution. Nevertheless, they put this possibility aside on the grounds that dust should be accelerated in the same way as the water-ice carrying particles and this was not indicated in the data. In section 3 we propose that if essentially all of the water-ice is released in the source regions of jet filamentary structures while the majority of the refractory dust is released from the inter-filament regions of the Type IIa jet complex, Protopapa *et al*'s observations can be explained.

Our analysis is based on a simplified analytic approach rather than a detailed computational modeling for it is not our intention to even attempt a simulation all of the detailed complexities of the inner coma (see section 2) but to only emphasize the physical aspects of our explanation of Protopapa *et al's*. (2014) observations, estimate mass flow of the responsible particulates, and connect the latter with the comet's hyperactivity. The linearity of the jet filaments implies that rotation of the nucleus has negligible effects for the particle size that is responsible for Protopapa *et al's* observed column densities. However, rotation becomes highly significant for the very largest, most weakly accelerated, particles (chunks) in the overall size distribution**.**

In section 2 we describe the data. In section 3 we outline a model for the structure of 103P's main jet ($J_1$), which is based on a concept for Type IIa jets (*i.e.* those driven by super-volatiles) developed in Belton (2010). We also provide an explanation of why the column densities of fine refractory dust and water-ice particles do not share the same radial distribution. In section 4 we apply our model to Protopapa *et al's* data. In section 5 we resolve the problem of hyperactivity by involving the dislodgement of large chunks of sub-surface material in the filament source regions. Finally, in section 6, we provide a brief discussion of our results and the limitations of the approach used.

## 2. 103P's coma, data limitations, and assumptions for the $J_1$ jet region.

There is little doubt that the inner particulate coma of 103P/Hartley 2 is a highly complex entity. The spatial distributions of water-ice bearing particulates and $CO_2$ are strongly correlated, while the distribution of $H_2O$ gas has quite a different character (Protopapa *et al.*, 2014). Dust is more evenly distributed around the nucleus than either $H_2O$ or $CO_2$. Within ~ 40 km of the nucleus the coma has not only a rapid expansion of water vapor accelerating fine dust outward from the surface (*e.g.,* Fougere *et al.*, 2013; Tenishev *et al.*, 2011), but has embedded within it a population of much larger, possibly centimeter-to-meter sized (the word size always means effective radius in this paper) particles as deduced from radar reflections (Harmon *et al.*, 2011) and EPOXI images (Kelley *et al.*, 2013, 2015; Hermalyn *et al.*, 2013). These coma particles show a relatively steep differential size-frequency distribution (SFD) with power law slopes ranging from -6.6 to -4.7 (Kelley *et al.*, 2013, 2015) and have speeds that hover near the escape speed from the nucleus (~ 0.3 ms$^{-1}$) to several 10 ms$^{-1}$ (Harmon *et al,* 2011; Hermalyn *et al.*, 2013). They have a velocity distribution that is far from radial with respect to the nucleus, that is somewhat weighted towards the anti-solar direction. This complexity may be amplified by sublimation of $H_2O$ from water-ice carrying aggregates in the coma, possibly leading to dis-aggregation, and thus adding further components to the coma.

The radar cross-section of this particle population is 0.89 km$^2$, some 15 times that of the nucleus itself (Harmon *et al.,* 2011), and shows radial speeds rising to ~30 ms$^{-1}$ (Harmon *et al.,* 2011). A rough calculation based on this cross-section suggests that the radar coma mass approaches ~$10^{-4}$ $M_{nucleus}$ ($M_{nucleus}$ = 4.3 x $10^{11}$ kg; Thomas *et al.,* 2013) assuming a bulk density of 535 kg.m$^{-3}$ (Preusker *et al.,* 2015)[3], a differential SFD slope of -4.7, and a minimum size of 0.02 m (Harmon *et al.,* 2011), This is considerably less than a revised estimate by Kelley *et al.* (2015) who finds a *visible* coma mass in the range 6 x $10^{-3}$ to 1.4 x $10^{-1}$ $M_{nucleus}$ assuming that the particles are dust/water-ice aggregates with the bulk properties of the nucleus.

Embedded in this coma are at least six jets ($J_1$ – $J_6$) emanating from the surface. Five of these show strong water-ice spectral signatures and these jets are also correlated with enhanced $CO_2$ outflows (Protopapa *et al.* (2014). Visual images (Syal *et al.,* 2013) show that the most prominent of these jets ($J_1$) consists of a complex of filamentary structures typical of the Type IIa jets described in Belton (2010).

*2.1 Spatial resolution of the observations.*

As noted by A'Hearn *et al.* (2011), a resolution of 10 to 12 m/pixel in the visible is insufficient to resolve the source region of the jet filaments. Their physical characteristics remain obscure. Since fine dust particles are distributed more evenly around the nucleus than are water-ice carrying particles, which are strongly coupled to the $CO_2$ distribution, we hypothesize that both water-ice carrying particles and a fraction of the observed dust are dislodged from the source regions at the base of the $CO_2$ jet filaments while the majority of the refractory dust is dislodged directly from the upper surface of the nucleus as a result of $H_2O$ sublimating through an ice-free mantle away from the filament source regions.

---

[3] In this work we consistently use this bulk density determination, which refers to comet 67P/Churyumov-Gerasimenko, and adjust mass calculations from other papers to this value.

Protopapa *et al's* (2014) data on column densities were obtained with the Deep Impact HRI-IR imaging spectrometer instrument at a spatial resolution of 55 m/pixel and then averaged with a 3 x 3 pixel box (Protopapa *et al*, 2014). The net resolution is apparently too low to separate out the individual filaments in the jets that are easily seen in the HRI-VIS images (Protopapa *et al.*, 2014; Figure 10) taken at approximately the same time at ~ 11 m/pixel and higher resolutions (Syal *et al.*, 2013). Thus the Protopapa *et al.* (2014) data represents a supposition of information over many filaments and inter-filament regions. In retrieving column densities, Protopapa *et al.* (2014) assumed that an optically thin situation applies. This assumption is consistent with the reported column densities that imply optical depths < 0.06, indicating that opacity is not an issue. We adopt this assumption in the calculations reported below.

*2.2 Radial distribution of water-ice and dust for the $J_1$ jet region.*

Protopapa *et al.* (2014) mapped the radial distribution of water-ice and dust in two jets ($J_1$ and $J_4$), however, because of the complex geometry associated with $J_4$ and the relatively straight forward geometry of $J_1$, we only consider the data for the latter in this analysis. The $J_1$ region also shows the strongest concentration of $CO_2$ and water-ice signatures. At time of observation the center-line of the jet structure was roughly propagating in the direction of the sun and is roughly orthogonal to the line of sight (Protopapa *et al.*, 2014).

Figure 1 shows column density profiles for dust and water-ice carrying particles in the comet's primary jet region $J_1$ transcribed from Protopapa *et al.* (2014; their figure 12). In the figure the average column density of the water-ice carrying particles has been increased by a factor of 100 for ease of display and the abscissa is measured from a center of curvature some distance, $\varrho'$, below the surface. Protopapa *et al.* (2014) estimate this distance to be ~ 935 m. The data is remarkable in that it represents material that is in close vicinity to the surface of the small lobe of the nucleus (effective radius, $R_s$ = 350 to 450 m) covering altitudes from ~ 440 to 1970 m (~ 1 to 5 $R_s$). As can be seen from plots of the predicted speeds of dust grains in various cometary environments (Figure 2), this is a range of altitude where the acceleration of dust by gases subliming from the surface is normally predicted to take place (Tenishev *et al.*, 2011). It is also a region in which the gas flow responsible for the acceleration of the particles is itself still accelerating. For example, in a simulation for 67P/Churyumov-Gerasimenko at a distance of 1.29 AU from the sun, Tenishev *et al.* (2008) estimated that this latter region can extend to an altitude of ~10,000 m on the sun-line of a typical Jupiter Family comet.

According to Protopapa *et al.* (2014) the run of column densities of dust with distance $\rho$ from the center of curvature shows a -0.99 ± 0.02 slope on a log-log plot. This is consistent with dust flowing at a constant, presumably terminal, speed. From this we deduce that the particles have low mass to area ratio, *i.e.,* the dust is very fine (say, ~ 1μ radius) and that its acceleration to terminal speed occurs rapidly almost entirely below an altitude of ~ 439 m (~ 1$R_s$).

In stark contrast, the column densities of water-ice particles show evidence of either increasing speed throughout the measured range of altitude or, possibly, the effects of another process, *e.g.*, sublimation, or, possibly, dis-aggregation. Protopapa *et al* (2014) assumed that this process was sublimation of icy particles (1μ effective radius) already at their terminal speed. They estimated this speed to be ~ 0.5 ms$^{-1}$ based on the lifetime of 1 micron pure icy grains near 1 AU (Hanner, 1981). However they noted that such a low terminal speed is in strong conflict with all existing predictions of the terminal speed of micron-sized dust particles.

*2.3 Size of the water-ice particles in the jet region.*

Protopapa *et al.* (2014) obtained and analyzed spectra of the particulate material in the region of the $J_1$ jet structure, which clearly shows the signature of water-ice (their Figures 7 - 9). They concluded that the shapes of the water-ice features were most consistent with a well separated mixture of dust and water-ice particles and that particles with effective radius < 5μ best fit the spectra. In their subsequent analysis Protopapa *et al.* assumed a particle size of 1μ. Here we will consider separately three possible sizes, 1μ, 3μ, and 5μ.

*2.4 Geometry of the $J_1$ jet.*

At the highest available spatial resolution, in the visible, the $J_1$ jet region is seen to be composed of a cluster of ~ 13 straight filamentary structures emanating from an area of ~ 0. 6 km$^2$ near the end of the small lobe (Syal *et al.* 2013). These structures populate a cone of half-angle ~ 27° on the sky and emerge roughly orthogonal to the curvature of the surface. This kind of jet structure, driven by $CO_2$, was classified as Type IIa in the taxonomic scheme devised in Belton (2010). The individual filaments are straight and highly collimated and, in this case, are made visible by a mixture of water-ice and dust particles that are entrained in the $CO_2$ gas flow. The filaments appear to be ≤ 100 m across at their base and they have a striking similarity to the filamentary jets seen in 1P/Halley (Keller *et al.*, 1987) and other comets whose filaments typically open with a full cone angle of ≈ 5 - 10°. We take 9° in these calculations to be consistent with the $CO_2$ jet model results in Belton (2010). We note that numerical experimentation shows that the value of this parameter has little effect on the fit to column densities. However, it is important in understanding the effects of rotation in the calculations since particles released at the base of the filaments and weakly accelerated along their length will eventually, due to rotation, leave the filament into the surrounding $H_2O$ outflow.

*2.5. The effects of rotation in the jets and the mean radial speed of the water-ice carrying particles.*

The tangential speed of the surface due to rotation (Belton *et al.,* 2013) in the vicinity of the $J_1$ jet region is ~ 0.14 m.s$^{-1}$. If the water-ice carrying particles were moving radially at an average speed as low as 0.5 m.s$^{-1}$ (Protopapa *et al.,* 2014) the effects of rotation should be noticeable at the region of observation (average radial distance 1210 m from the surface) for the angular displacement of the particles from the instantaneous radial direction would be ~ 17°. We find no obvious evidence in the visible images for such a separation. The straightness of the filaments implies that the particles that define them are moving with much higher speed by the time they reach the region of observation. For example, if the filaments showed a displacement of < 2° from the instantaneous radial direction they would have an average speed to the altitude of the observations > 4.5 m.s$^{-1}$. We shall find (section 4.2) that the particles have a typical speeds of ~1 to 10 m.s$^{-1}$ in the region of observation.

The particles dislodged at the base of the filaments will have a wide range of sizes. The very largest of these (chunks), which are just lifted from the sub-surface by the $CO_2$ flow, will be distinguished by an extremely weak radial acceleration and, as a result, Coriolis force will prevent them from participating in the flow directly along the linear jet filaments over the distance to the region of observation. This can be illustrated by a simple geometric argument: Consider one of the largest chunks that is dislodged at the base of a filament and that can move independently of the surface. The source region of the filament is at a distance R from the axis of rotation. It will be essentially neutrally buoyant and, in inertial space, its trajectory will be a straight path that takes it away from the source region that is fixed on the surface of the rotating nucleus. The direction in which the chunk moves will be also be determined by the rotation. After one quarter of the rotation period (~4.5 h; Belton *et al.,* 2013) the chunk will have travelled a distance $(1/2)\pi R$ and be moving parallel to the axis of the jet. It will be separated from the axis by the distance R. A simple geometric construction shows that the angle between the jet axis and the direction from the source region to the chunk is θ =arctan(2/(π-2)), or θ = 60.3°. For the chunk to remain in the jet filament, the filament would have to have a full cone angle of 120.6°, far larger than the 5 – 10° that is observed. Positively buoyant chunks of lesser size that were dislodged at the same time, including the fine particulates that define the filament, will fill part of the volume of space between this boundary and the filament axis. In the real world however, the particle dynamics will be considerably more complicated as the chunks leave and enter $CO_2$ rich flows and enter surrounding $H_2O$ flows with different momentum transport capabilities while passing through different regions of the gravitational field of a nucleus with complex shape. The buoyancy of the chunks can be expected to vary relatively rapidly on time-scale less than the rotation period as they move through the inner coma. Any attempt to follow the detailed trajectories of individual clunks with the goal of connecting the source of chunks with the particles observed by Kelley *et al.* (2013, 2015) and Hermalyn *et al.* (2013) would be a formidable task and is not attempted in this paper.

*2.6. Physical parameters for the $J_1$ jet region.*

The water-ice carrying chunks can be thought of as mini-nuclei with properties similar to the nucleus itself. For the size of the largest chunks released we take the maximum size of coma particles (assumed to have the properties of "mini-nuclei") found by Kelley et al. (2015), *i.e.,* 4m effective radius. The chunks are assumed to have the same sublimation properties as the nucleus surface, although as "fresh" objects from the sub-surface they could have higher sublimation rates. We have already noted the bulk density of d = 535 ± 35 kg m$^{-3}$ (Preusker *et al.,* 2015)

We also require information on the levels and uncertainties in the observed fluxes of $H_2O$ and $CO_2$. Since there are at least five jet regions producing water-ice particles and $J_1$ is clearly very strong, the water ice injected into the coma by $J_1$ should be a fraction, w, of the total amount injected. For the present calculations we take *w* = 0.5 while recognizing that this, of necessity, is an arbitrary choice. Examination of Figure 10 in Protopapa *et al.* (2014) indicates that this may be a minimum value for w which will affect our estimate of the contribution of $J_1$ to the degree of hyperactivity as noted in section 4.2 below. The rates at which water is injected into the coma from other sources (*e.g.,* sublimation from the nucleus waist region, sublimation from the rest of the nucleus, and directly by water-ice particles) have been estimated by Fougere *et al.* (2013). They find, from an assessment of the results of five independent observers, that at the time of the observations between ~150 kg s$^{-1}$ and ~360 kg s$^{-1}$ of water is injected into the coma as water-ice particles with 210 kg s$^{-1}$ as a nominal value. Accepting these results and the value of w noted above, our estimate of the rate at which $J_1$ injects water ice into the coma has a nominal value of ~105 kg s$^{-1}$. For $CO_2$ we again use the results of Fougere *et al.* (2013) who find from modeling the EPOXI IR spectrum that the sub-solar jet (*i.e.*, $J_1$) was producing ~9.2 10$^{26}$ s$^{-1}$ (~67 kg s$^{-1}$) of $CO_2$. Fougere *et al.* give no uncertainty with this estimate, but we would not be surprised if it was not at least ± 10%. Finally, we need an estimate of the terminal speed of $CO_2$ along the jet filaments. As there is, to our knowledge, no observational measurement of this this quantity we use the results of calculations by Belton (2010) as a guide, *i.e.*, ~ 400 m s$^{-1}$.

## 3. A simplified model for the Hartley 2 $CO_2$ jet filaments.

We adopt the model proposed by Belton (2010) to describe the basic physics governing the jet structure. In this model the filaments are narrowly collimated stream-tubes that act as conduits for $CO_2$ emanating from a cluster of isolated source regions on the surface. It is supposed that the filaments maintain their straightness (during daylight hours) as the result of a pressure balance between the flow of super-volatiles within them and the ambient flow of a sublimating water atmosphere. They are made visible by particles released in the source regions and then entrained in the gas flow. During the night the ambient $H_2O$ atmosphere collapses and the filamentary structure is lost. However, during this period super-volatiles may continue to be released at the surface and simply expand into a vacuum in a broad flow (continuing to carry water-ice carrying particles and fine dust with it) in the manner first described by Kitamura (1986, 1987). Such behavior, *i.e.,* emission of $CO_2$ from the $J_1$ region on the night-side, may have already been observed by Feaga *et al.* (2014). The flow interior to a filament, dominated by $CO_2$, also contains a small admixture of sublimating water vapor. The mixing of these two components at the base of the filament is presumed to result in a turbulent region at the surface where a mixture of both icy chunks and refractory and water-ice particulates are dislodged into the accelerating gas. Exterior to, and between, the filaments water is sublimated through an ice-free mantle to release small dust grains directly from the surface (*e.g.,* Gombosi *et al.*, 1985).

      Because the size of the individual filament source areas are uncertain the surface area between filaments Is uncertain by a factor of ~2 – 40. Nevertheless it is larger than the area within the filaments themselves and the observed radial variation column density of the dust over the jet region is expected to be dominated by the flow between filaments. In contrast the run of column density of icy material is dominated by the flow within the filaments. It is this separation of domains, as well a possible difference in size between the two types of particles, that we propose as the reason for the difference in the radial dependence of dust and water-ice column densities seen in the measurements of Protopapa *et al.* (2014). The two types of particles are accelerated in quite different gas flows (Figure 3). Other models of jets have been proposed, *e.g.,* Combi *et al.* (2014), Yelle *et al.* (2004), Syal *et al.* (2013), and Crifo *et al.* (2004) for earlier work, but they have not emphasized, this separation between dust and water-ice dominated regions.

*3.1 Column density of water-ice particles in the $J_1$ jet region.*

For the purpose of simplifying this calculation we assume with Protopapa *et al*. (2014) that the water-ice particles are all of the same physical size with effective radius, *a* (in our calculations we consider *a* = 1, 3, and 5μ). While this is obviously a severe limitation to the calculation we feel that it is necessary at this stage since we don't know the dependence of the slope of the differential size-distribution of the particles with altitude within the jet filaments. As noted above, Kelley *et al.* (2013, 2015) have estimated the power law slope of the differential size-frequency distribution (SFD; index ranging from – 6.6 to - 4.7) for the larger particles (≥ 1 cm) in the coma. However, because of effects on particle size like sublimation and de-aggregation processes and dynamical considerations it is far from clear how to relate the SFD for the large particles found by Kelley *et al.* to the SFD of much smaller micron-sized particles in the jet filaments or even to the overall SFD of particles as they are released at the source of the jets. It is because of this complexity that we have followed Protopapa *et al.* (2014) in representing the SFD of particles in the jet filaments by particles of a single size that is intended to represent the dominant contributor to Protopapa *et al's* observations. We recognize that this will have an effect on our numerical results for the mass flow and mean speed of particles along jet filament. A rough indication of the magnitude of these effects can be obtained by comparing the results of our calculations for the range of particle sizes (1 – 5 μ) that Protopapa *et al.* (2014) find are the most significant contributors to the to the light scattered by the jet. We find uncertainties of a factor of 2 or 3 are possible (Section 4.2).

We visualize a filament $f$ as a collimated stream-tube with circular cross-section that expands along its length at a small half-angle $\theta_f$. *i.e.,* solid angle $2\pi(1 - cos\theta)$ (see Figure A1). The mid-line of the filament cone is inclined to the plane of the sky at an angle $\phi_f$. The number density of icy particles at a distance $l$ from the center of curvature is $n_f(l)$ particles.m$^{-3}$ where (Appendix A):

$$n_f(l) = \frac{q_f}{l^2 V_f(l)} \tag{1}$$

The particles flow along the filament with speed $V_f(l)$ m.s$^{-1}$. The effective area at the base of is $2\pi\rho'^2(1 - cos\theta_f)$ m$^2$ and in this area fine water-ice particles are dislodged at a rate $q_f$ s$^{-1}$ster$^{-1}$. The total rate at which fine water-ice particles are produced in the filament is $Q_f = 2\pi q_f(1 - cos\theta_f) \, s^{-1}$.

The column densities derived by Protopapa *et al.* (2014) are based on a circular arc of width $\delta\rho$ at distance $\rho$ from the center of curvature in the plane of the sky centered on $J_1$ (Figure A1(a)). The arc subtends an angle $\psi$ on the sky, which defines the cone within which a bundle of *m* filaments resides and has an area on the sky of $A = \rho\psi\delta\rho$. Protopapa *et al's* column densities, $N(\rho)$ are the sum of the contributions of all of the filaments within the arc averaged over this area. Equation A11, as worked out in Appendix A, gives an expression for column density in terms of the production rate of particles at the base of the jet, the speed profile of the particles, and the geometric circumstances. While Eq. A11 should give a fair representation of the distribution of column densities within the jet region, it is, nevertheless, too complex to be useful for application to the data at hand. For example, we have little information on the spatial distribution of the individual filaments other than the angular limits of the jet region. The distribution of their degrees of collimation or the distribution of their individual strengths are also unknown. However, we do know that the data are averages over the angular extent $\psi$ (~ 54°) of the jet region and that the light is scattered in the optically thin limit. Moreover, we have a fair idea of the collimation of filaments from other comets, *i.e.*, $\theta$ ~ 9° (Keller *et al.*, 1987; Yelle *et al.*, 2004) and the geometric situation suggests that the average value of $\emptyset_f$ is near 0°. If we assume that the bundle of actual filaments can be replaced by the super-position of *m* optically thin, identical, filaments with average parameters as noted above, we can write the following approximations:

$$N(\rho) = \frac{Q}{\pi\psi\rho(1-cos\theta)} \int_0^\vartheta d\alpha \int_{y_l(\vartheta,\alpha)}^{y_u(\theta,\alpha)} \frac{dy}{(1+y^2)V(y,\rho)} \qquad (2)$$

where

$$y_u(\theta,\alpha) = \tan(\beta(\theta,\alpha)) \qquad (3)$$

and

$$y_l(\theta,\alpha) = -tan(\beta(\theta,\alpha)) \qquad (4)$$

*3.2 The shape of the speed function $V(y,\varrho)$.*

To calculate $V(y,\rho)$ a full gas-dynamic calculation would normally be required (*e.g.*, as in Tenishev *et al.*, 2008 or Combi *et al.*, 2012 where the DSMC method is used), however the detailed properties of the $CO_2$ gas-flow are not well enough understood and this puts such a calculation well beyond the scope of this investigation. It is for this reason that we choose to represent $V(y, \rho)$ in a parametric form. To determine the shape of this function we use a number of detailed calculations of the acceleration of dust grains of various radii in various coma environments that have been published (Tenishev *et al,* 2011; Combi *et al.* 2012, Gombosi *et al.* 1985 and earlier studies that are referenced in the review by Wallis (1982)). The later studies include, as a matter of course, treatments of the acceleration of the coma gas itself and the changing drag coefficient.

By numerical experimentation and visual fitting we find that in all cases at altitudes >100 m a function of the form:

$$V(h) = V_0 \left( exp\left(-\frac{L}{\mu+h}\right) + A \right) \qquad (5)$$

emulates the calculated speeds (Figure 2). Here $V_t = (1 + A)V_0$ is the terminal speed of the grains whose actual value ultimately depends on where the particles become detached from the accelerating gas. *h* is the altitude, and L, μ, and A are constant parameters that characterize the particular coma environment. Table 1 lists typical values of these parameters for the various dust acceleration curves shown in Figure 2. From a physical standpoint L is the characteristic length over which the overall acceleration takes place and μ is a length which determines (with the value of A) the initial speed in the immediate vicinity of the source region. A characteristic of Eq.5 is that the acceleration is always positive, slowing off as the terminal speed, $V_t$, is approached at high altitudes. This behavior explains the (observed) steeper slope of the water-ice curve in Figure 1 and its (as yet unobserved) transition at higher altitudes (on a log-log plot) to ρ$^{-1}$ slope. It does not explain, however, the precise curvature that is observed at lower altitudes in the data. Numerical experimentation shows with visual fitting that this behavior can be induced by increasing the speed at the lowest altitudes near the surface, which is the purpose of the constant parameter, A. Inserting this function into Eq.(7) we obtain:

$$N(\rho) = \frac{(Q/V_t)(1+A)}{\pi \psi \rho (1-\cos\theta)} \int_0^\vartheta d\alpha \int_{y_l(\vartheta,\alpha)}^{y_u(\theta,\alpha)} \frac{dy}{(1+y^2)(\exp(-L/(\mu+\left(\varrho\sqrt{(1+y^2)}-\varrho'\right)+A)} \qquad (6)$$

Inspection of this equation shows that as $\varrho \to \infty$, $N \to Const/\rho$, as seen, for example, in the dust column density profile, and that as $\varrho \to \rho'$, *i.e.,* towards to the surface, $N$ tends to a constant value dependent on the total particle production rate. Fitting $N(\rho)$ to the Protopapa *et al.* (2014) data yields estimates for $\mu, L, A,$ and $Q/V_t$.

## 4. Application of the jet model to the $J_1$ jet region data.

*4.1 Speed and mass flow in the region of observation.*

Figure 1 shows our fit to the water-ice data. This was done by numerical experimentation and visual fitting. While we judge it to be quite satisfactory it is certainly not perfect. This is at least partly due to the use of Eq. 5 for the speed function whose shape is merely an acceptable approximation (*cf.* Figure 2) to the real (*i.e.,* modeled) speed function in certain environments. The derived parameters are: $\mu$ = 550±50 m, L = 8800±1000 m, A = 0.0031±0.002, and $Q/V_t$ = 6.5 (±2) X $10^9$ $m^{-1}$. The uncertainties in the above parameters are not formal uncertainties, rather they have been estimated by a trial and error visual fitting procedure. In Figure 4 we show the speed function $V(\rho)/V_t$ for these parameters and compare it with a calculated profile for 1μ dust particles in a typical $H_2O$ coma. Clearly, the acceleration of the water-ice carrying particles in the jet filaments is much lower than is typical for small refractory dust in the inter-filament regions.

To relate the column densities found by Protopapa *et al.* (2014), who assumed 1μ for the particle radius, to particles of radius *a* we have $Q_a = 9 \times 10^{-13} Q / \alpha a^2$ (Since Protopapa *et al.* do not specify the albedo of the water-ice particles they assumed in their paper we have presumed a value of 0.9. Also, following Protopapa *et al.* we have ignored possible but unknown differences in phase function that are not expected to be large). With the values of $d$ = 1000 kg.$m^{-3}$, α = 1, and $Q/V_t$ from the above fit to Protopapa *et al*'s column densities, we find

$$V(1200\ m) = 9.6 \binom{+8.5}{-5.4} 10^{-3}\ V_t \quad \text{(m.s}^{-1}) \tag{7}$$

$$\dot{M}_a = m_a Q_a = 24.5\ (\pm 7.5) a V_t \quad \text{(kg.s}^{-1}) \tag{8}$$

for the speed, *V*, and the mass flow, $\dot{M}_a$, of fine water-ice particles in the region of observation (altitude ≈ 1200 m) in terms of the terminal speed of the particles. Here $m_a$ is the mass of a particle of size *a*. The large range of uncertainty takes account of the full range of uncertainty in the various contributing parameters.

*4.2 The terminal speed, $V_t$, and the speed and mass-flow through the region of observation.*

We have little concrete knowledge about terminal speeds for fine water-ice particles, only that they are surely less than the terminal speed, $u_\infty$, of the $CO_2$ gas in the filamentary structure that is accelerating them, *i.e.,* $u_\infty$ ~ 400 m.$s^{-1}$ (Belton, 2010). The radar observations of Harmon *et al.* (2011), which are sensitive to particles as small as ~.02 m, show radial speeds in the coma skirt up to ~30 m.s$^{-1}$. We presume that this high speed is at, or close to, the terminal speed of the smallest particles, 0.02 m, detectable by the radar. Using this and an approximate relationship developed by Wallis (1982), based on a constant gas speed theory, we can make a rough attempt to extrapolate the radar information to the terminal speed of the fine water-ice particles observed by Protopapa *et al.* (2014). Wallis (1982) finds:

$$\frac{V_t}{u_\infty} = 1/[A' + B'a^{0.5}] \qquad (9)$$

Where $u_\infty$ is the terminal speed of the accelerating gas, *A'* is a constant that reflects the drag of the particles on the gas, and B' is a constant measuring the coupling of the particles to the gas. As the drag on the gas due to the particles tends to zero, *A'* → 0.9. In our calculation we take *A'* = 1 and $u_\infty$ = 400 m.s$^{-1}$ as representative of the physical conditions in the $CO_2$ filament flow. Applying the radar result ($a = 0.02m$; $V_t = 30\ m.s^{-1}$; $u_\infty = 400\ m.s^{-1}$) to Eq. 9 gives B' = 8.7 x 10$^1$ m$^{-0.5}$. Applying this result to the fine water-ice particles observed by Protopapa et al. (2014) gives: $V_t$ = 367, 347, 335 ms$^{-1}$ for a = 1µ, 3µ and 5µ respectively. Applying this range of terminal speeds to Eq. 7 we find the particles accelerating rapidly through the region of observation, but still at low speed relative to $V_t$ (see Figure 4). The particles enter the region of observation at speeds near ~1 m.s$^{-1}$ and leave at ~ 10 m.s$^{-1}$. A typical speed in the middle of the region is 3 m.s$^{-1}$.

The mass flow (Eq. 8) through the J$_1$ region for the range of particle size we are considering lies between 3.3 x 10$^{-2}$ and 5.4 x 10$^{-2}$ kg.s$^{-1}$ with a nominal value of ~ 2 x 10$^{-2}$ kg.s$^{-1}$. Given our choice of w in section 2.6, **t**his is ~ 0.04% of what is needed to support the observed degree of hyperactivity (section 2.6) in the comet. If w is > 0.5 the contribution of the fine water-ice particle flowing through the filamentary structure of J$_1$ is even less. It is also ~40 times the maximum mass-flow estimated by Protopapa *et al.* (2014) with a model based on the sublimation of water-ice particles in the jets. Clearly, In order to find the cause of the hyperactivity we must look to the larger chunks that are dislodged at the base of the filaments, but which, due to nucleus rotation, do not flow up the filaments into the region of Protopapa *et al's* observations,

### 5. Hyperactivity in comet 103P/Hartley 2.

*5.1 Sublimation of H$_2$O in the coma relative to the nucleus.*

As noted in the introduction, Kelley *et al.* (2015) have basically resolved the problem of the hyperactivity of 103P. If the particles that they observe can be thought of as "mini-103P comets" they present, according to Kelley *et al.* (2015) a total cross-section of 0.2 – 2 km$^2$. Considering them to be spheres this represents a total area of ~0.4 – 4 km$^2$ to sunlight in the coma that is sublimating H$_2$O at rates/unit area that are presumably similar to that of the nucleus surface, *i.e.,* they have the same "active fractional areas" (If the mini-comets are in some sense "fresh" they may sublimate at a higher rate, but this is impossible to determine directly from the observations). The total area of the nucleus is 5.24 km$^2$ (Thomas *et al.* 2013) of which roughly half should at any time be actively sublimating H$_2$O, *i.e.*, ~ 2.6 km$^2$. This is well within the range of sublimating area found for the coma particles. Thus the coma particles, viewed in this way, can double or even triple the water production rate expected for the comet nucleus itself. Typical active fractions for typical comet nuclei are near ~0.1, so the coma particles seen by Kelley *et al.* (2015) could increase this to as high as ~0.3. Groussin *et al.* (2004) found, utilizing the Infrared Space Observatory, values of 0.3 and ~1 for 103P's active fraction depending on heliocentric distance.

> *5.2 Production rate of the large coma particles and implication for type IIa jet source regions.*

Based on the estimated speeds of large coma particles by Hermalyn *et al.* (2013), most of which are well above the escape speed of ~0.3 m.s$^{-1}$, Kelley *et al.* (2015) estimate**,** using a rough order of magnitude escape speed of 0.1 m.s$^{-1}$, that the dynamical lifetime of large coma particles is ~ 60 h and from this they deduce a mass loss rate from the coma of ~1.7 x 10$^3$ kg.s$^{-1}$ (here we have adjusted their estimate of mass loss from an assumed bulk density of 300 kg.m$^3$ to 535 kg.m$^3$ to be consistent with our assumptions). They also note that this loss rate could possibly be an order of magnitude higher if the relevant escape speed was comparable with the radar velocity dispersion (4 m.s$^{-1}$, Harmon et al., 2011). In fact, Harmon et al. have provided their own estimate of the coma loss rate for large particles based on a "simple grain ejection model" (see Harmon et al., 2004, for the model). They find a much lower loss rate of 300 kg.s$^{-1.}$ They are not specific about the bulk density that they used in this calculation except that a density near 1000 kg.m$^{-3}$ is appropriate for the top few meters and that they have consistently used 500 kg.m$^{-3}$ in their previous modelling for coma mass loss rates (Harmon et al., 2004). We assume that 500 kg.m$^{-3}$ is appropriate for their calculation, which is essentially the same as the value of 535 kg.m-3 used throughout this paper. Unfortunately, Harmon et al. (2011) did not have the benefit of the particle size distribution measured by Kelley et (2013), which will affect the results of their calculation. Clearly there is clearly great uncertainty in these estimates ranging over an order of magnitude. In the nominal calculations that follow we will accept Kelley et al's (2015) particulate loss rate of 1.7 x 10$^3$ kg.s$^{-1}$ since it employs a fair estimate of the speeds of particles within a reasonably well defined region of the coma centered on the nucleus. Nevertheless, we recognize the considerable, but exceptionally difficult to estimate, uncertainty that must be associated with this value.

First, consider the largest particles found by Kelley *et al.,* which have an effective radius of 4 m and a mass of ~1.4 x $10^5$ kg. This, together with the above mass loss rate, gives the shortest average time between the injection of such particles into the coma from the source region of a filament. If the $J_1$ jet region produces half of the material injected into the coma, then the gas flowing through each of its ~13 filaments can dislocate such a large particle (chunk) in a time no shorter than ~5.6 hr. The actual time between the injection of such big chunks is almost certainly much longer but, on average, shorter than the 60 hr. noted above. Thus a maximum of one of the largest chunks per one to three rotation periods would seem plausible.

Second, Kelley *et al's* mass loss rate should be roughly equivalent to the excavation rate in the source region of the filaments. If, as may be the case (Belton, 2010), the diameter of the source region at the base of a filament is in the range 50 – 100 m and the excavation is cylindrical with vertical sides, the depth of the excavation over 30 days around perihelion would reach some 44 – 171 m. This may well be an overestimate because it assumes that the flow in these jets is constant around perihelion passage, which may not be true. It is interesting that on 67P Vincent *et al.* (2015) have discovered 18 quasi-circular, straight-sided, pits, some 10 – 100 m in diameter that tend to cluster in small groups (several also show signs of current activity), with depths of at least 100 ~ 200 m. Vincent *et al.* suggest that these features may be sinkholes in the surface; here we advocate that some of these pits could be features that in the past supported, now exhausted, type IIa jet outflows given the similarity of the depth of excavation numbers and diameters reported above.

## 6. Discussion

With the exception that the chunks need not be pure water-ice, we have shown that A'Hearn *et al's* (2011) hypothesis on the source of 103P's hyperactivity is valid if the chunks are dislodged from the subsurface in the source regions of Type IIa, $CO_2$ driven, jets. The assumption that the chunks have the same bulk properties of the nucleus itself appears to be just adequate to explain the degree of hyperactivity displayed by the comet. However, this result could be improved if the newly dislodged chunks, owing to lack of previous exposure to sunlight, sublimate $H_2O$ with higher active fraction than the general surface of the nucleus. We make considerable use of the results of Kelley *et al.* (2015) that provide overall constraints on the sublimation area of the chunks and the mass-flow through the inner coma. These constraints not only lead to the result on hyperactivity noted above but also on the depths to which jet source regions are excavated during perihelion passage (~100 m). While there is no direct observational evidence from 103P on the detailed morphology of its jet source regions, we are of the opinion that, at least, some of the active pits groupings discovered on 67P by Vincent *et al.* (2015) that have diameters and depths consistent with our calculations, may well be, now essentially exhausted, source vents of super-volatiles from the interior.

The primary focus in this investigation was to provide explanations for the run of column densities of water ice and dust particles with radial distance observed by Protopapa *et al.* (2014) in the $J_1$ jet complex and why they are so different. The difference we ascribe to the dust originating directly from the nucleus surface between individual jet sources, while the origin of the water-ice particles is confined within the sources of the jet filaments. Thus most of the dust is accelerated by $H_2O$ directly from the surface, while the water-ice is accelerated by slower moving $CO_2$ (Belton, 2010) within the filamentary structure of the jets.

Our parameterization of the speed and acceleration of small particles with altitude, which is based on detailed dynamical calculations of particles of various sizes in two different comets, allowed us to compute the speed and acceleration of the water particles through the region of Protopapa *et al's* region of observation and the mass-flow in these particles. The typical speed at ~3 m.s$^{-1}$ is almost ten times higher and the mass-flow ~40 times higher than in a competing model based on the sublimation of ice particles as they flow through the jets by Protopapa *et al.* (2014). However, our initial expectation that the mass-flow in the filaments could explain the hyperactivity fell short when it was realized that the large chunks needed to explain the hyperactivity would be accelerated so slowly that they would be quickly deviated out of the jet structure by Coriolis forces.

As noted in section 2.5 the trajectories of the chunks as they make their way into and through the inner coma and eventually escape (or crash on the surface) is not considered in this paper partly because of the complexity of the calculations (*e.g.* Byram *et al.*, 2007) and partly because the lack of detailed observational information that is needed to constrain them. However, we note that Hermalyn *et al.* (2013) and Kelley *et al.* (2013) have begun such an investigation by measuring the velocities of chunks in the inner coma, characterizing their spatial distribution about the nucleus, and investigating the effects of solar radiation pressure and sublimation of their dynamics.

Our treatment provides only a rough explanation of both hyperactivity and Protopapa et al's (2014) observations for the calculations that we have made are clearly very crude and their theoretical basis contains gross simplifications. Chief among these are the assumptions of a single particle size flowing through the filament structure, the assumption of Belton's (2010) unproven Type IIa jet model, and, finally, the assumption that we can calibrate the terminal velocities of fine particles in the jets by reference to much larger particles observed by radar. Other problems include the assumption that the filaments in the J₁ jet complex are all equal in strength, and that their spatial distribution can be averaged to be roughly orthogonal to the direction of the line of sight. While we are of the opinion that these assumptions are highly significant, we are, nevertheless, of the opinion that none are lethal to our physical explanation of the observed radial distribution of water-ice particulates and why it differs from that of refractory dust or even to the general magnitude of the numerical results.

**Acknowledgements.** Belton Space Exploration Initiatives, LLC, provided support for M.J.S. Belton, who acknowledges the assistance and important contributions of Dr. S. Protopapa in providing a detailed review and analysis of an early draft of the paper. The author also acknowledges the help of Dr. M. S. P. Kelley for advance notice about corrections to his 2013 paper. The Tucson cometary lunch group also provided some insightful and critical suggestions.

**Appendix A: Evaluating column densities**

The column densities derived by Protopapa *et al.* (2014) are based on a circular arc in the plane of the sky of width $\delta\rho$ at distance $\rho$ from a center of curvature that is defined by the cone that contains the jet filaments (Figure A1(a)). The arc subtends an angle $\psi$ which defines the cone within which a bundle of $f$ filaments resides. The arc has an area $\rho\psi\delta\rho$. Protopapa's column densities, $N(\rho)$, are the average of the sum of the contributions from *m* filaments distributed over the area of the arc:

$$N(\rho) = \frac{1}{\rho\psi\delta\rho} \sum_{f=1}^{f=m} N_f(\rho, \phi_f, \theta_f) \quad , \tag{A1}$$

We visualize a filament as a narrow cone with half-angle $\theta_f$, *i.e.,* solid angle $2\pi(1-cos\theta_f)$, whose center line is inclined to the sky plane by angle $\phi_f$ (Figure A1(b)). $N_f$ is the total number of icy particles in filament $f$ that lie within the arc.

The computation of $N_f$ requires a double integration, one in the plane, P, containing the center line of the filament and orthogonal to the line-of-sight (LOS), and a second along the LOS in the region of arc occupied by filament $f$. Writing $N_{xf}(\rho, \phi_f, \theta_f, \alpha)$ for the column density along the line-of-sight $x(\alpha)$ where $\alpha$ is the angle in the plane P between the centerline of the filament and a line within the filament from the center of curvature to the point of intersection of the line-of-sight $x(\alpha)$ with P we have (Figure A1(c)):

$$N_f(\rho, \phi_f, \theta_f) = 2\rho\delta\rho \int_0^{\vartheta_f} \sec(\gamma(\varphi_f, \alpha)). N_{xf}(\rho, \phi_f, \theta_f, \alpha) d\alpha \qquad (A2)$$

Which gives

$$N(\rho) = \frac{2}{\psi} \sum_{f=1}^{f=m} \int_0^{\vartheta_f} \sec(\gamma(\varphi_f, \alpha)). N_{xf}(\rho, \phi_f, \theta_f, \alpha) d\alpha \qquad (A3)$$

To compute $N_{xf}(\rho, \phi_f, \theta_f, \alpha)$ we require an expression for the number density of particles $n_f(l)$ within the filament at a distance $l$ from the center of curvature. Writing $q_f$ for the rate per steradian at which fine water-ice particles are injected at the base of the filament and $V_f(l)$ as their speed, consideration of the flux of particles flowing through a small element of the filament, $dl$, at $l$ yields:

$$n_f(l) = \frac{q_f}{l^2 V_f(l)} \qquad (A4)$$

The number of particles injected into the filament per second is:

$$Q_f = 2\pi q_f (1 - \cos\theta_f) \qquad (A5)$$

Integration along the line-of-sight $x(\alpha)$ where $l^2 = \rho^2 + x(\alpha)^2$, i.e., the origin of x is taken as the point where the LOS crosses the plane of the sky, yields with a change of variable to $y = \frac{x}{\rho}$:

$$N_{xf}(\rho, \phi_f, \theta_f \alpha) = \frac{Q_f}{2\pi\rho(1-\cos\theta_f)} \int_{y_l(\phi_f,\theta_f,\alpha)}^{y_u(\phi_f,\theta_f,\alpha)} \frac{dy}{(1+y^2)V_f(y,\rho)} \qquad (A6)$$

Consideration of the geometry of the filament and the LOS (Figure A1(c)) shows that the integration limits are:

$$y_u(\phi_f, \theta_f, \alpha) = \tan(\gamma(\phi_f, \alpha) + \beta(\theta_f, \alpha)) \qquad (A7)$$

and

$$y_l(\phi_f, \theta_f, \alpha) = \tan(\gamma(\phi_f, \alpha) - \beta(\theta_f, \alpha)) \qquad (A8)$$

where

$$\gamma(\phi_f, \alpha) = acos(sin^2\alpha + cos^2\alpha . cos\varphi_f) \tag{A9}$$

and

$$\beta(\theta_f, \alpha) = acos(\frac{cos\theta_f}{cos\alpha}) \tag{A10}$$

These expressions give:

$$N(\rho) = \frac{1}{\psi\pi\rho} \sum_{f=1}^{f=m} \frac{Q_f}{(1-cos\theta_f)} \int_0^{\theta_f} \sec(\gamma(\varphi_f, \alpha)) \left( \int_{y_l(\phi_f,\theta_f,\alpha)}^{y_u(\phi_f,\theta_f,\alpha)} \frac{dy}{(1+y^2)V_f(y,\rho)} \right) d\alpha$$

$$\tag{A11}$$

**Tables**

| Environment | $V_t$ (ms$^{-1}$) | µ (m) | L (m) | a (microns) | Ref. |
|---|---|---|---|---|---|
| Jet simulation for 67P/C-G | 388 | 250 | 450 | 0.1 | 1 |
|  | 184 | 170 | 470 | 1 | 1 |
|  | 66 | 130 | 500 | 10 | 1 |
| Daytime coma simulation for 67P/C-G | 275 | 580 | 870 | 0.1 | 2 |
|  | 109 | 550 | 970 | 1 | 2 |
|  | 35 | 550 | 1200 | 10 | 2 |
| 1P/Halley | 262 | 135 | 240 | 0.5 | 3 |

**Table 1.** Typical parameters for the speed function $V = V_t(\exp(-L/(\mu + (\rho - \rho')) + A)$ from fitting theoretical predictions of the acceleration of dust particles of radius a in various coma environments (see also Figure 2). A=0 in all of these model fits.

**Figure captions**

Figure 1. Average column density profiles for dust and water ice particles in the $J_1$ jet region transcribed from Proptopapa *et al.* (2014; their figure 12). The abscissa $\rho$ is distance on the sky from the center of curvature of the $J_1$ jet region some 935 m below the surface of the nucleus. For a detailed description of the nature of the data see the text (section 2). The thick solid line is the fit of our model (section 4) to the water-ice data. The thin solid line is the fit to dust data by Protopapa *et al.* (2014). Note the water-ice column densities are increased by a factor of 100 above their actual value for display purposes.

Figure 2. Predicted speeds of dust particles (•) in the inner parts of various coma environments as a function of altitude $h$ above the surface of the nucleus. These values have been digitized from the original papers noted below. Super-posed are fitted curves of $V = V_0(\exp(-L/(\mu + (\rho - \rho'))) + A)$. *Panel* A: From Combi *et al.* (2012) for 1 micron particles in the context of a jet feature (L=490 m, μ=200 m, $V_t$ =186 ms$^{-1}$); *Panel* B: From Combi *et al.* (2012) for 10 micron particles in a jet feature (L=500 m, μ=130 m, $V_t$ = 66 ms$^{-1}$); *Panel* C: From Combi *et al.* (2012) for 1 micron particles in an expanding coma with no jet calculated by Tenishev *et al.* (2011) (L=970 m, μ=550 m, $V_t$ =109 ms$^{-1}$); *Panel* D: From Gombosi *et al.* (1985) for 0.5 micron particles in the innermost coma of a comet near 1AU (L=240 m, μ=135 m, $V_t$ =262 ms$^{-1}$). In all cases and at altitudes > 100 m the function with A=0 is found to give an excellent account of the predicted speeds and their rates of change.

Figure 3. *(Dotted line).* Ratio of momentum transport in a filament ($CO_2$) versus that in the inter-filament region ($H_2O$) as a function of altitude according to the model in Belton (2010). *(Solid line).* Speed in the $H_2O$ inter-filament region. *(Dot-dash line).* Speed vs. altitude in the $CO_2$ filament.

Figure 4. Comparison of relative speeds of 1μ dust (*dashed* line) predicted for a typical coma environment (67P/Churyumov-Gerasimenko at 1.29 AU; Combi *et al*, 2012) and that deduced for dusty water-ice particles (*solid line*) in the filaments of the jet structure $J_1$ in comet 103P/Hartley 2 at 1.06 AU.

Figure A1. Visualization of the geometry of a filament and viewing direction. The three panels define the various angles employed in Appendix A. (a) View of the plane of the sky and Protopapa *et al's* (2014) measurement arc at a distance ρ from the center of curvature. The arc extends over an angle $\psi$ and is $\delta\varrho$ wide. The hatched region represents the contribution of particles in the arc by a filament. (b) Geometry of a filament showing the angles $\alpha, \varphi,$ and $\theta$. (c) Geometry of a filament showing the angles $\alpha, \gamma, \beta$ and the x-direction integration limits $y_u$ and $y_i$.

**Figure 1**

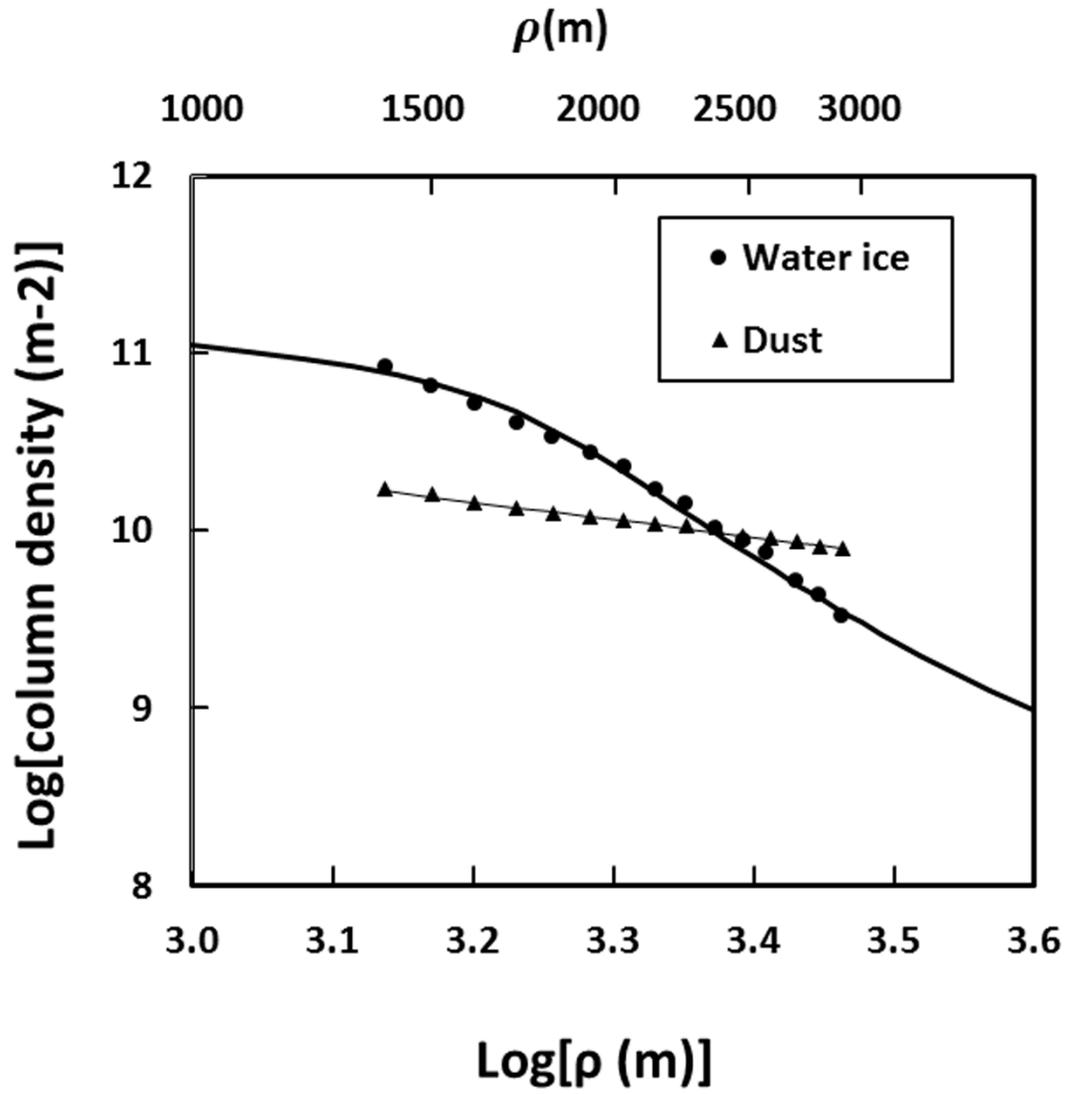

**Figure 2**

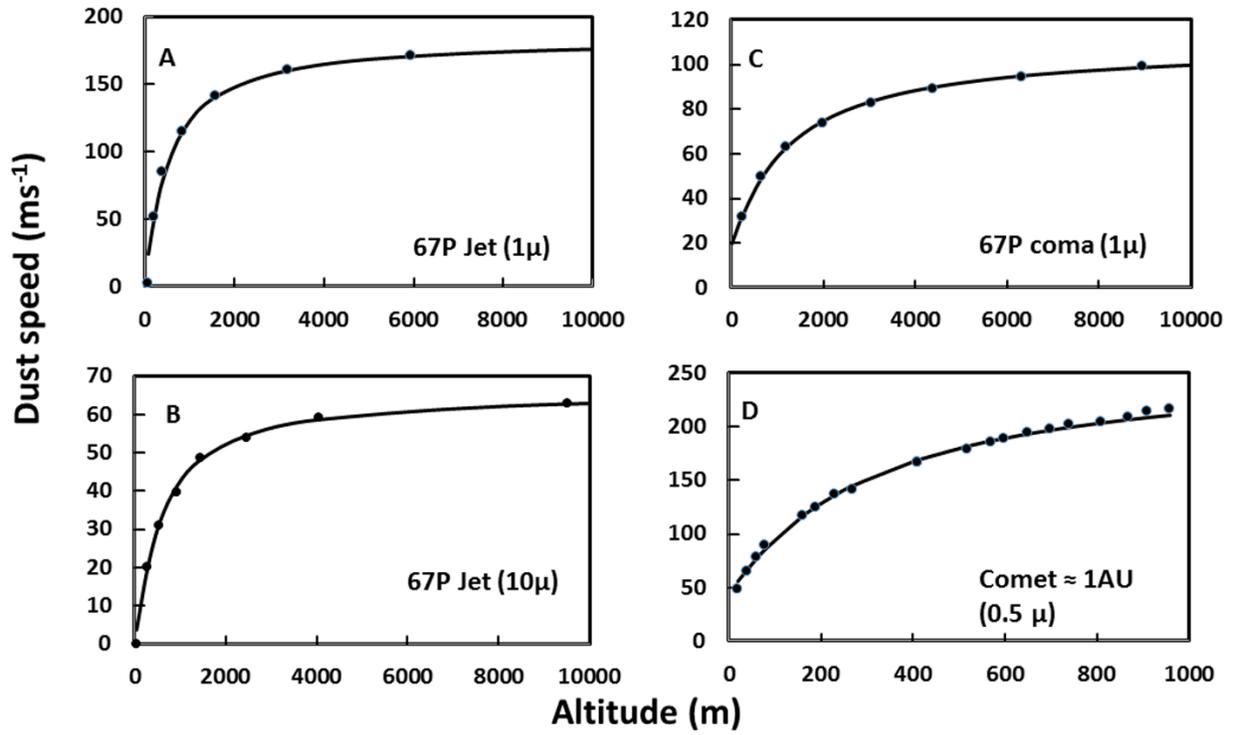

**Figure 3**

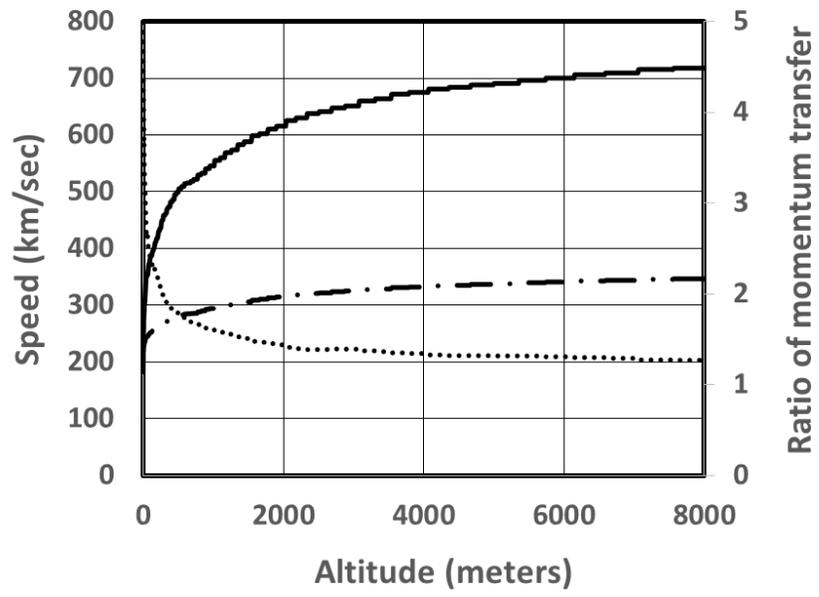

**Figure 4**

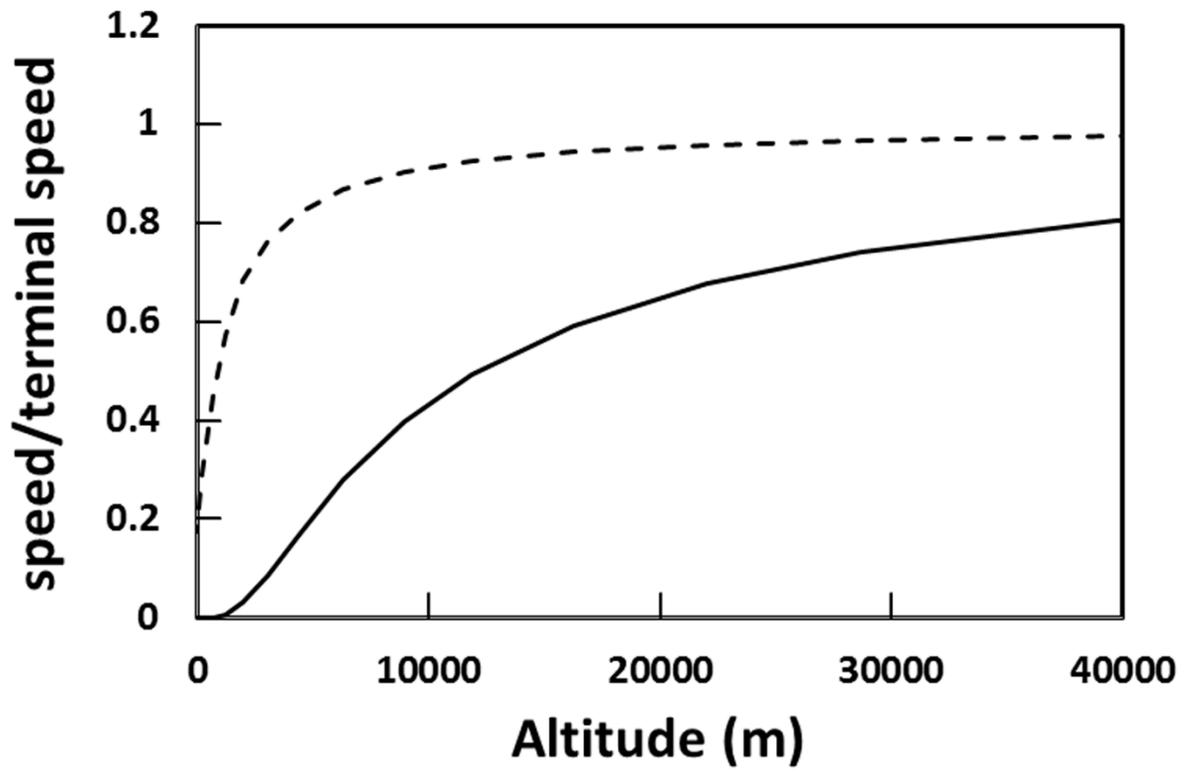

**Figure A1**

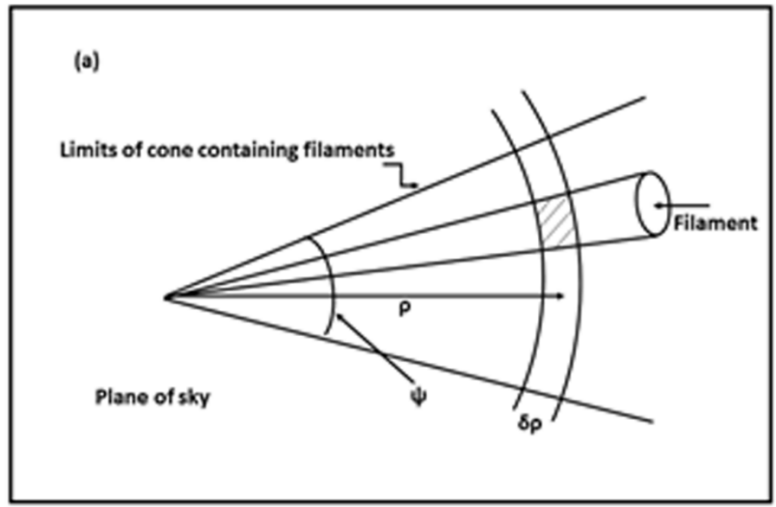

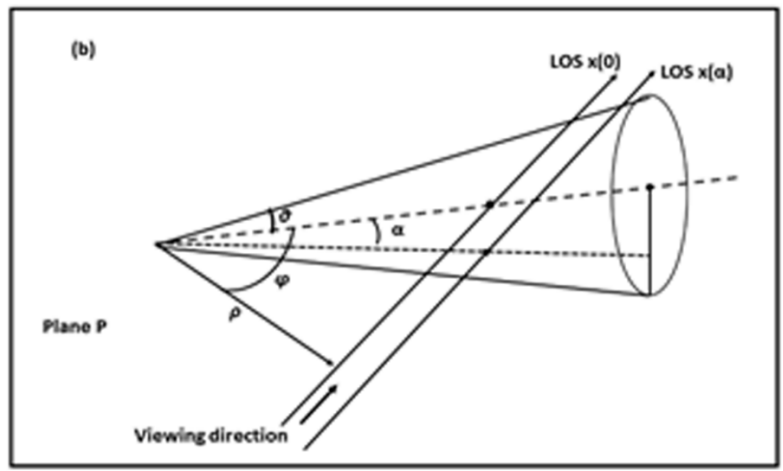

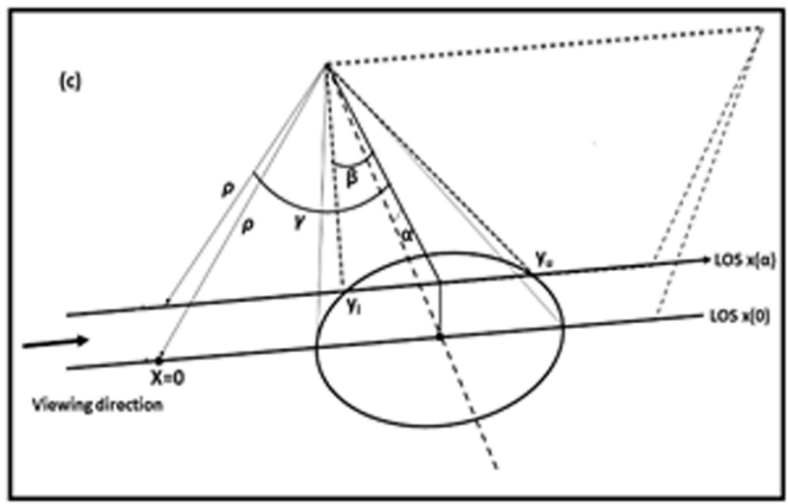